\documentclass[doublecol]{per}
\LogoHeight{1.75cm}

\usepackage[T1]{fontenc}
\usepackage[utf8x]{inputenc}
\usepackage{graphicx}
\usepackage[utf8x]{inputenc}
\usepackage{siunitx}
\usepackage{lmodern}
\usepackage{pifont}
\usepackage{amsmath}
\usepackage{amssymb}
\usepackage{xcolor}

\newcommand{\footnoteremember}[2]{%
\footnote{#2}%
\newcounter{#1}%
\setcounter{#1}{\value{footnote}}%
}

\title{Energy cascade and the four-fifths law in superfluid turbulence}

\author{J. Salort\inst{1} \and B. Chabaud\inst{1} \and
  E. Lévêque\inst{2} \and P.-E. Roche\inst{1}} 

\institute{%
  \inst{1} Institut Néel, CNRS/UJF - BP 166, F-38042 Grenoble cedex 9, France, EU\\
  \inst{2} Laboratoire de Physique de l'ENS de  Lyon, CNRS/Université Lyon, F-69364 Lyon cedex 7, France, EU} \shortauthor{J. Salort,
  B. Chabaud, E. Lévêque and P.-E. Roche}

\pacs{47.37.+q}{Hydrodynamic aspects of superfluidity; quantum fluids}
\pacs{67.57.De}{Superflow and hydrodynamics}
\pacs{67.25.dk}{Vortices and turbulence}

\abstract{\color{black} The 4/5-law of turbulence, which characterizes
  the energy cascade from large to small-sized eddies at high Reynolds
  numbers in classical fluids, is verified experimentally in a
  superfluid $^4\mathrm{He}$ wind tunnel, operated down to
  \SI{1.56}{K} and up to $R_{\lambda}\approx 1640$. The result is
  corroborated by high-resolution simulations of Landau-Tisza's
  two-fluid model down to \SI{1.15}{K}, corresponding to a residual
  normal fluid concentration below \SI{3}{\percent} but with a lower
  Reynolds number of order $R_{\lambda}\approx 100$. Although the
  Kármán-Howarth equation (including a viscous term) is not valid
  \emph{a priori} in a superfluid, it is found that it provides an
  empirical description of the deviation from the ideal 4/5-law at small scales and
  allows us to identify an effective viscosity for the superfluid,
  whose value matches the kinematic viscosity of the normal fluid regardless
  of its concentration.}
\begin{document}

\maketitle


\section{Introduction}

At low temperature, but above the so-called \emph{lambda} transition, liquid $^4{\mathrm{He}}$ is a classical
fluid known as He~I. Like air or water, its dynamics obeys the Navier-Stokes equation.  When such a fluid is
strongly stirred, its response is dominated by the non-linearity
 of the Navier-Stokes equation. The dynamics of such a system,
known as ``turbulence'', was first pictured by Richardson in 1920 and
theorized by Kolmogorov in 1941\cite{kolmogorov1941}.  The kinetic energy,
injected at some large scales, cascades down across the so-called
\emph{inertial} scales until it reaches the dissipative scales.  It can be
derived from the Navier-Stokes equation that this energy flux across
scales results in skewed distributions for the velocity
increments. This prediction (the only exact result known for
turbulence) is sometimes referred to as the Kolmogorov's 4/5-law.  It is
recalled later in this paper.

When liquid $^4\mathrm{He}$ is cooled below $T_{\lambda}\approx\SI{2.17}{K}$ (at
saturated vapor pressure), it
undergoes the lambda phase transition. The new phase, called He~II, can be
described within the so-called \emph{two-fluid model} \cite{landau1941}, \emph{i.e.} the superposition of a viscous ``normal fluid'' 
and an inviscid ``superfluid'' with quantized vorticity, these two components being coupled by a
mutual friction term. The fraction 
$\rho_s/\rho_n$ --- where $\rho_s$ and $\rho_n$ are respectively the
densities of the superfluid and normal components --- varies with
temperature from $0$, at $T_{\lambda}$, to $\infty$ in the
zero-temperature limit.  When He~II is strongly stirred, a tangle of
quantum vortices is generated. This type of turbulent flow is
characterized as ``{quantum turbulence}'' or ''{superfluid
turbulence}``. For an introduction to quantum turbulence, one may refer
to \cite{vinen2002,sergeev2011}.

The focus of this letter is on intense turbulence in He~II at finite
temperature, \emph{i.e.} $T_{\lambda} > T \geq \SI{1}{K}$. In such conditions,
most of the superfluid kinetic energy distributes itself between the
mechanical-forcing scale (at $\sim$\SI{1}{cm} in \cite{roche2007}) and
the inter-vortex scale (at $\sim$\SI{4}{\micro\meter} in
\cite{roche2007}). Excitations at smaller scales are strongly damped
by the viscosity of the normal component \cite{vinen2002}.  At scales
larger than the inter-vortex spacing the details of individual
vortices are smoothed out (``continuous'' or ``coarse-grained''
description) and superfluid turbulence can be investigated with the
same statistical tools as classical turbulence. An important open
question is how superfluid turbulence compares with classical
turbulence. Experimental studies have revealed differences regarding vorticity spectra \cite{roche2007,bradley2008} but also striking
similarities concerning decay-rate scaling
\cite{stalp1999,niemela2005,chagovets2007,walmsley2008}, drag force
\cite{rousset1994,smith1999,fuzier2001} and $k^{-5/3}$ scaling for the energy spectrum
\cite{maurer1998,salort2010}.  This latter is consistent with the
existence of an energy cascade (as described by Kolmogorov's theory), however no direct proof has
been reported yet, as stressed recently during the \emph{Quantum Turbulence
Workshop} in Abu Dhabi \cite{benzi2011} (see also the conclusion of
\cite{samuels1999}).

The main goal of this paper is to test in superfluid turbulence the characteristic 4/5-law
of the energy cascade.  \color{black} To account for
departure from the ideal 4/5-law at small scales, the classical
Kármán-Howarth equation is assessed. \color{black} As a side result, it is
showed that the superfluid inherits viscosity from the normal component
even when the normal fraction is very low, \color{black} therefore making the velocity signal 
of a superfluid (obtained by an inertial anemometer like a Pitot tube)  hardly 
distinguishable from the one of a classical fluid.\color{black} We consider
 experimental velocity fluctuations measurements obtained in a
\SI{1}{m}-long cryogenic helium wind tunnel at high Reynolds number, as
well as results from direct numerical simulations of the continuous
two-fluid model, at lower Reynolds numbers but fully resolved down to 
the mean inter-vortex scale.

\section{Local velocity measurements}

\begin{figure}[htbp]
\begin{center}
\includegraphics[scale=.4]{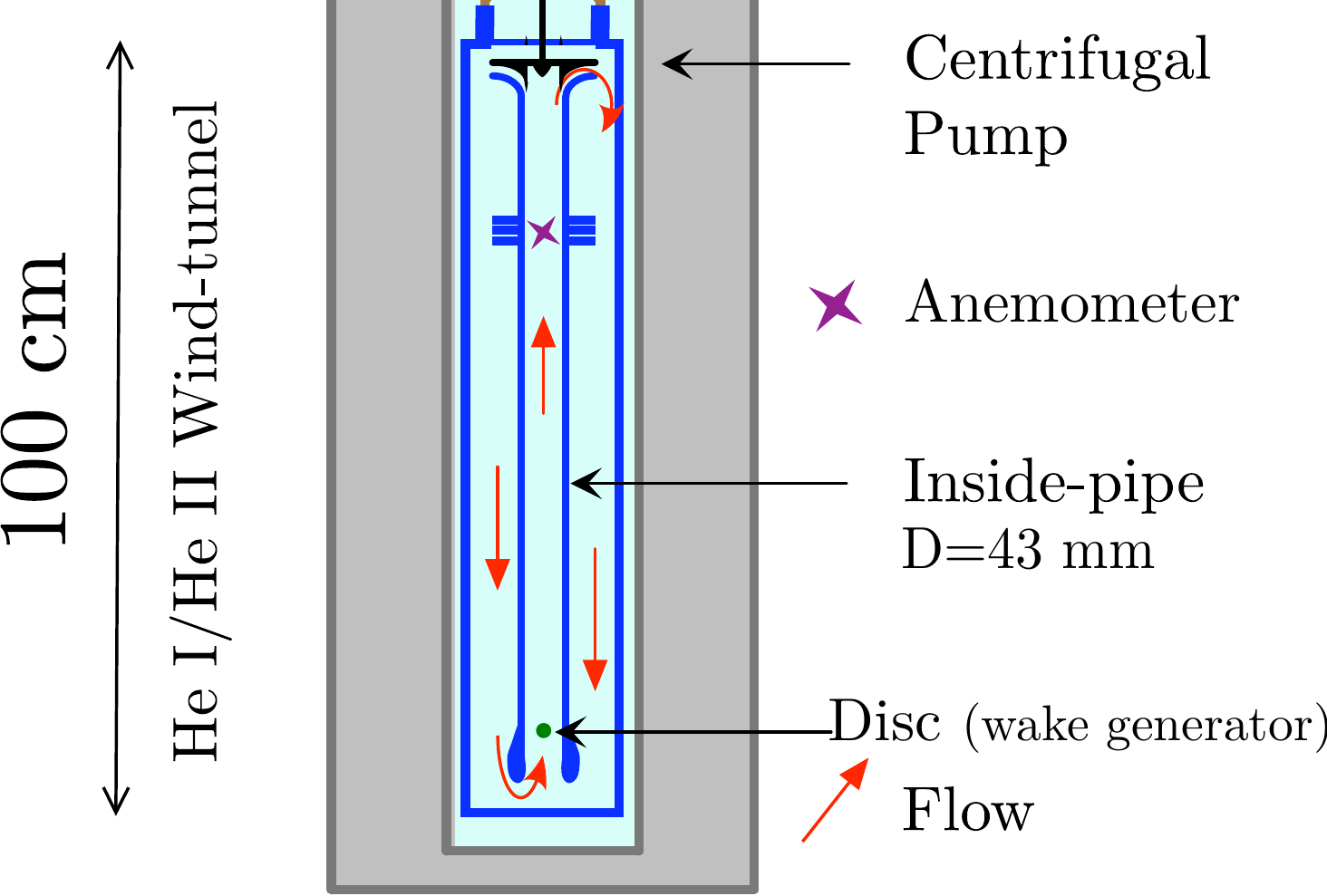}
\end{center}
\vspace*{-5mm}
\caption{Wind tunnel (in blue) in the cryostat (in
  gray)}\label{fig:Toupie}
\end{figure}

\begin{figure}[htbp]
\begin{center}
\includegraphics[scale=.90]{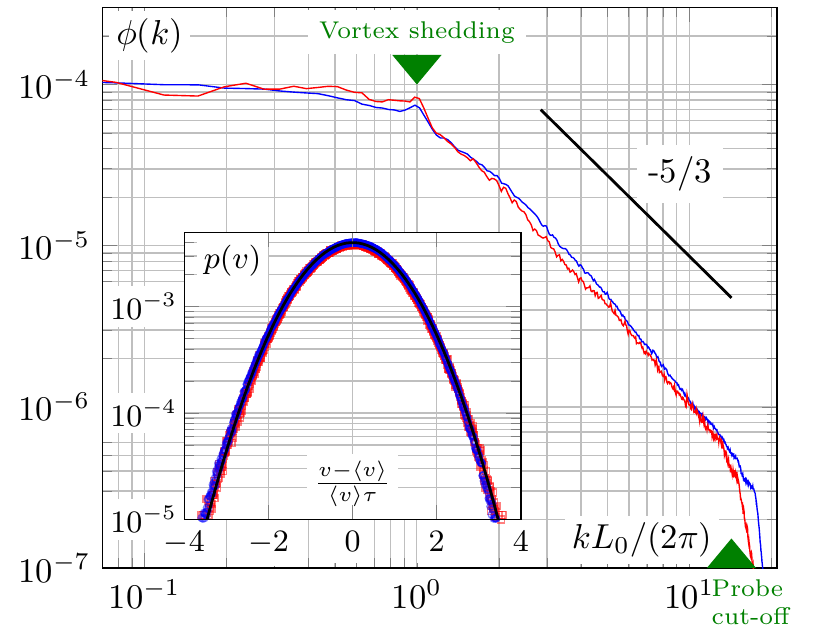}
\end{center}
\vspace*{-5mm}
\caption{Experimental 1D velocity power spectrum above and below the
  superfluid transition. Red line: $T=\SI{2.2}{K}>T_{\lambda}$ at
  $R_{\lambda}\approx 1640$. Blue line:
  $T=\SI{1.56}{K}<T_{\lambda}$. Inset: Velocity probability density
  distribution above and below the superfluid transition. Black line:
  Gaussian distribution.}\label{fig:R5Spectre}
\end{figure}

Local velocity measurements have been performed in the far wake of a disc
in the wind tunnel sketched in figure \ref{fig:Toupie}. The disc
diameter $\varnothing_d$ is half the pipe diameter. The probe, located
downstream at $x/\varnothing_d \approx 21$, was operated both above
and below the superfluid transition, down to \SI{1.56}{K} for which
$\rho_s/\rho_n \approx 5.8$. The wind
tunnel is pressurized by more than \SI{1}{m} of static liquid to
prevent cavitation.
The turbulence intensity, $\tau = {\sqrt{\left< (v(t)-\left<v\right>)^2 \right>}}/{\left< v \right>}$,
where $v(t)$ is the local flow velocity and $\left< . \right>$ stands
for time average, is close to \SI{4.8}{\percent}; \color{black} the mean
velocity is $\left< v \right> = \SI{1}{m/s}$. \color{black} The
forcing length scale, $L_0$, is obtained from the frequency of vortex
shedding: $f_0=\left<v\right>/L_0$. This latter is estimated from the velocity
spectrum (see figure \ref{fig:R5Spectre}).  The typical Strouhal number
\begin{equation}
  St = \frac{f_0\varnothing_d}{\left< v\right>}  = \frac{\varnothing_d}{L_0}
\end{equation}
is found close to 0.35  both above and below the superfluid
transition.  At $T=\SI{2.2}{K}$, where liquid helium is a
classical  fluid   with kinematic viscosity 
$\nu=\SI{1.78e-8}{m^2/s}$ \cite{donnelly1998}, the Reynolds
number 
$Re = v_{\rm rms}L_0/\nu = \num{1.8e5}$.
The  Reynolds number based on Taylor microscale is here approximated by $R_{\lambda} = \sqrt{15~ Re} \approx 1640$.

The local anemometer is the probe labeled as \ding{172} in
\cite{salort2010}. It is based on a stagnation pressure measurement
(miniature ``Pitot tube'' probe).  It measures the pressure overhead
resulting from the stagnation point at the tip of the probe,
 which is pointing upflow.  Above the
 superfluid  transition, the measured pressure
$p_{\text{meas}}(t)$ is
\begin{equation}
p_{\text{meas}}(t) = p(t) + \frac{1}{2}\rho v^2 \label{eq:pmeasclassical}
\end{equation}

Following \cite{maurer1998}, a similar
expression can be found for the measured pressure below the lambda
transition using the continuous two-fluid description of He~II:
\begin{equation}
p_{\text{meas}}(t) = p(t) + \frac{1}{2}\rho_n v_n^2 + \frac{1}{2}\rho_s v_s^2
\end{equation}
where $v_n$ is the velocity of the normal component and $v_s$ is the
velocity of the superfluid component.  Yet, physically, the probe is
sensitive to the flux of momentum on its tip.  It is therefore
convenient \cite{kivotides2002} to rewrite the measured pressure in terms of the ``{momentum
velocity}'', $\vec{v}_m$, with
\begin{equation}
\rho\vec{v}_m = \rho_n\vec{v}_n+\rho_s\vec{v}_s
\end{equation}
where $\rho = \rho_n+\rho_s$. This leads to
\begin{equation}
p_{\text{meas}}(t) = p(t) + \frac{1}{2}\rho v_m^2 + \frac{\rho_n\rho_s}{2\rho}\left(v_n-v_s\right)^2 \label{eq:pmeas}
\end{equation}

This equation is similar to the one standing in classical fluid (Eq.\
\ref{eq:pmeasclassical}) except for an additional term. It has been argued
theoretically\cite{vinen2002} and shown numerically\cite{salort2011}
that, in the fully-developed turbulent regime, the normal and superfluid components
are nearly locked at inertial scales.  Therefore, $\left( v_n - v_s
\right)^2 \ll v_m^2$ and since $\rho_n\rho_s \leq \rho^2$, the last
term in Eq.\ \ref{eq:pmeas} can be neglected\footnote{When the
  turbulence intensity $\tau$ is small, the same approximation is obtained
  with the weaker hypothesis: $\left< v_s \right> = \left< v_n
  \right>$, \emph{i.e.} the normal and superfluid components are locked at large
  scales \cite{kivotides2002}. The additional term is of order
  $\tau^2$ at most, and can be neglected.}.

The calibrations of the probe above and below the superfluid
transition are consistent  with each other 
within \SI{10}{\percent}. Discrepancies come mainly from experimental
uncertainties. In practice, the calibration obtained below
$T_{\lambda}$, where the signal is cleaner, was used to determine the
mean values obtained in normal fluid.
A numerical 4\textsuperscript{th}-order Butterworth low-pass filter is
applied to the velocity time series to suppress the probe organ-pipe
resonance \cite{salort2010}.  The filtered velocity time series
are converted  into 
spatial  signals  using the instantaneous Taylor's
frozen turbulence hypothesis\cite{pinton1994}, \emph{i.e.} 
the velocity at location $x$ is mapped to the velocity at time $t$, so that 
\begin{equation}
v(x)=v(t) \mbox{ with } x = \int_0^tv(\tau)\mathrm{d}\tau
\end{equation}
Velocity power
spectra and probability distribution functions (PDF) are estimated from
 velocity series recast in space, $v(x)$, and shown in figure
\ref{fig:R5Spectre}.
As expected, power spectra exhibit a Kolmogorov's
$k^{-5/3}$ scaling and the velocity PDF is nearly
Gaussian. The spectra above and below the superfluid transition are found
nearly identical. The wave number are normalized by the forcing scale
$L_0$ (see above). Let us mention that the observed cut-off at high $k$ results from
the finite resolution of the probes and not from a dissipative
effect.

The longitudinal velocity increments,  here along the streamwise
direction,  are defined as
\begin{equation}
\delta v(x;r) = v(x+r)-v(x)
\end{equation}

\begin{figure}[htbp]
\begin{center}
\includegraphics[scale=.90]{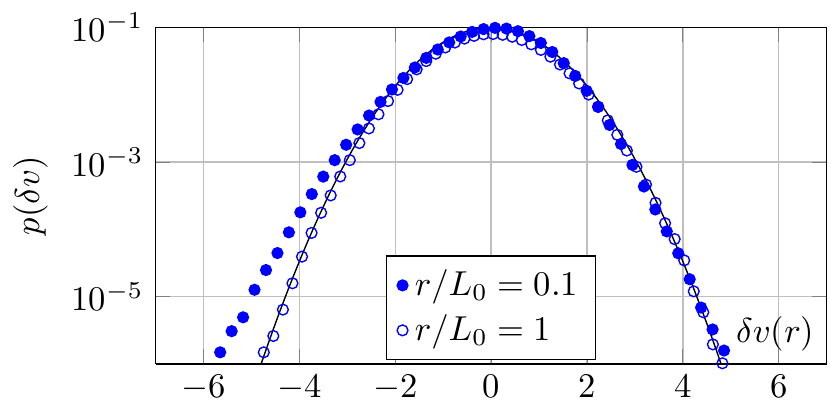}
\end{center}
\vspace*{-5mm}
\caption{Experimental histogram of the longitudinal velocity
  increments at large and intermediate scales in a superfluid
  turbulent flow ($T=\SI{1.56}{K}$). Solid black line: Gaussian
  PDF.} \label{fig:histoIncr}
\end{figure}

\begin{figure}[htbp]
\begin{center}
\includegraphics[scale=.90]{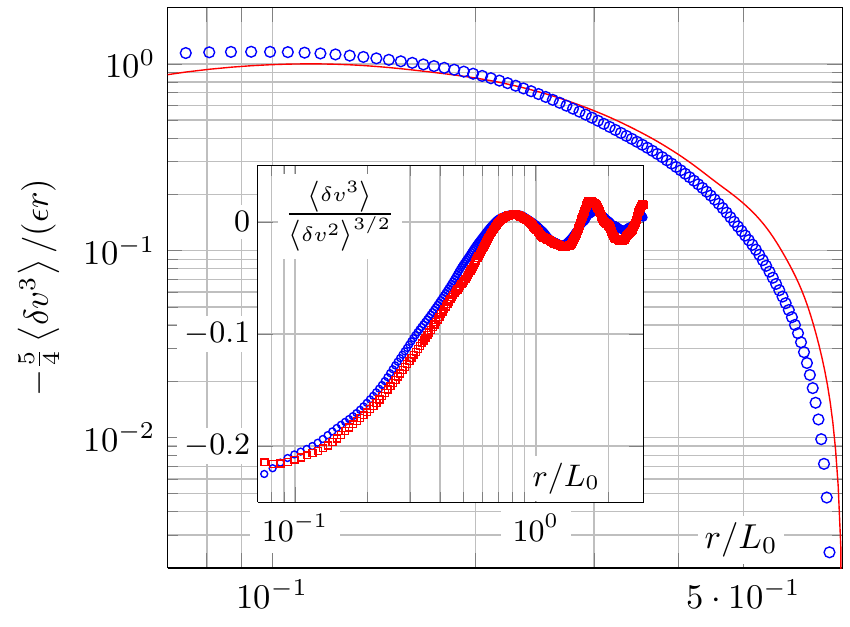}
\end{center}
\vspace*{-5mm}
\caption{Experimental third-order velocity structure function
  compensated by the 4/5-law (Eq.\ \ref{eq:epsilonKolmogorov})
  obtained in superfluid helium at $T=\SI{1.56}{K}$
  (blue circles) and in classical liquid helium at $T=\SI{2.2}{K}$
  (red squares). Inset: Skewness of the distribution of
  longitudinal velocity increments (same color code).  The
  smallest abscissa $r/L_0 = \num{7e-2}$ corresponds to the probe
  cut-off. The oscillation at large scales is related to the frequency of the vortex shedding.}\label{fig:howarth}
\end{figure}

The PDF of $\delta v(x;r)$ for a given separation $r$ is shown in figure
\ref{fig:histoIncr}. It is fairly Gaussian at large scale ($r\approx L_0$) and
clearly skewed on the negative side at smaller scales
($r \approx L_0/10$). The skewness $S(r)$ is
 defined as
\begin{equation}
S(r) = \frac{\left< \delta v(r)^3\right>}{\left< \delta v(r)^2 \right>^{3/2}}
\end{equation}
where $\left< . \right>$ stands for  space
 average.  $S(r)$ is shown in the inset of
figure \ref{fig:howarth}. 

Above the superfluid transition, $S(r)$ is known to be linked
to the transfer rate (or flux) of the energy cascade\cite{monin1971}.
Its value at our smallest resolved scale is fairly compatible
with the values (close to \num{-0.23}) already reported in the literature (in the limit of vanishing scale $r$); a review of
experimental values for $208 \leq R_{\lambda} \leq 2500$ may be found in
\cite{chevillard2006}.  The negative sign of $S(r)$ is a direct evidence that energy cascades from large to small scales.

Below the superfluid transition, the value of the skewness is found
nearly identical to the value  above the
superfluid transition. This is a strong hint that energy cascades
in a similar fashion above and below the superfluid transition.
More quantitatively, in classical \color{black} homogeneous and
isotropic \color{black} turbulence, 
the 4/5-law states that \color{black}
\begin{equation}
\color{black}
\left< \delta v(r)^3 \right> = -\frac{4}{5}\epsilon r \label{eq:epsilonKolmogorov}
\end{equation}
where $\epsilon$ stands for the mean dissipation rate of kinetic energy. 
This equation is only valid for inertial scales $r$, at which cascade dynamics prevails. It
is often cited as the only exact result of classical fully-developed
turbulence\color{black}, \emph{i.e.} for asymptotically large
$Re$\color{black}. It is our motivation to test its validity in quantum
turbulence. \color{black} In our experimental setting
$R_{\lambda} \approx 1640$ and, therefore, Eq.\
\ref{eq:epsilonKolmogorov} is expected to be ``approached'' in a finite inertial range of scales \cite{antonia2006}. \color{black}

In order to compare superfluid experimental data to this classical prediction,
$\epsilon$ needs to be estimated at first. Getting an accurate estimate of $\epsilon$ from
experimental data is not trivial. A \color{black} common \color{black} practice is to use
the third-order structure function and assume the 4/5-law. This method is known to yield
reasonable estimates of $\epsilon$ for $R_{\lambda} \gtrsim
1000$\cite{moisy1999,ishihara2009}. \color{black} Since our aim is here to
assess the 4/5-law, we can not use this method directly.
However, previous experiments have shown that $\epsilon$ does not change
when the superfluid transition is crossed (keeping the same mean-flow velocity above and below the transition) \cite{salort2010}.
Therefore, we have estimated $\epsilon$ from the
4/5-law using He~I velocity recordings --- where it is known to
hold, since He~I is a classical fluid --- and then used that estimate to
compensate the third-order velocity structure function obtained in
He~II.  We have obtained $\epsilon = \SI{5.4(3)e-3}{m^2/s^3}$. \color{black}

\color{black} We observe a ``plateau'' for nearly half a decade of
scales, corresponding to the resolved inertial range of the turbulent
cascade (see figure \ref{fig:howarth}). \color{black} The value of this
``plateau'' is comparable above and below the superfluid transition,
within an experimental uncertainty of about \SI{25}{\percent}.  This may be viewed as
the first experimental evidence that the 4/5-law (Eq.\
\ref{eq:epsilonKolmogorov}) remains valid in superfluid turbulence, at
least at the largest inertial scales. 

\section{Direct numerical simulations}

In this section, we examine turbulent velocity fields obtained from a pseudo-spectral simulation of He~II dynamics
in a cubic box (with resolution $512^3$--$1024^3$ and periodic boundary conditions). 
Stationarity is ensured by an isotropic external force acting at some large scale  $L_0$. 
The numerical procedure is detailed in \cite{salort2011}. The
dynamical equations write as
\begin{equation}
\frac{D\vec{v}_n}{Dt} = -\frac{1}{\rho_n}\nabla p_n + \frac{\rho_s}{\rho}\vec{F}_{ns} 
+ \frac{\mu}{\rho_n}\nabla^2\vec{v}_n+\vec{f}_n^{ext} \label{eq:vn}
\end{equation}
\begin{equation}
\frac{D\vec{v}_s}{Dt} = -\frac{1}{\rho_s}\nabla p_s - \frac{\rho_n}{\rho}\vec{F}_{ns}+\vec{f}_s^{ext} \label{eq:vs}
\end{equation}
where indices $n$ and $s$ refer to the normal and superfluid components,
respectively; $\vec{f}_n^{ext}$ and $\vec{f}_s^{ext}$ are external (divergence-free) forces; $\mu$ is the dynamic viscosity. The mutual coupling
force is approximated by its first-order expression:
\begin{equation}
\vec{F}_{ns} = -\frac{B}{2}\left|\vec{\omega}_s\right|\left( \vec{v}_n-\vec{v}_s \right)
\end{equation}
where $\vec{\omega}_s=\nabla\times \vec{v}_s$ is the superfluid vorticity and
$B= 2$ is taken as the mutual friction coefficient \cite{barenghi1983}. 
The normal and superfluid velocity fields are assumed incompressible, \emph{i.e.} $\nabla\cdot \vec{v}_s = \nabla \cdot \vec{v}_n = 0$. 

In our simulations, we fix the cut-off resolution 
at the value of the mean quantum inter-vortex distance, $\delta$. This latter is estimated from
the quantum of circulation, $\kappa$, around a single superfluid vortex
and from the average vorticity,
\begin{equation}
\delta^2  = \frac{\kappa}{\sqrt{\left<\left|\vec{\omega}_s\right|^2\right>}}
\end{equation}
This truncation procedure was validated by the accurate prediction of
the vortex line density in experiments \cite{salort2011}.

\begin{figure}[htbp]
\begin{center}
\includegraphics[scale=.90]{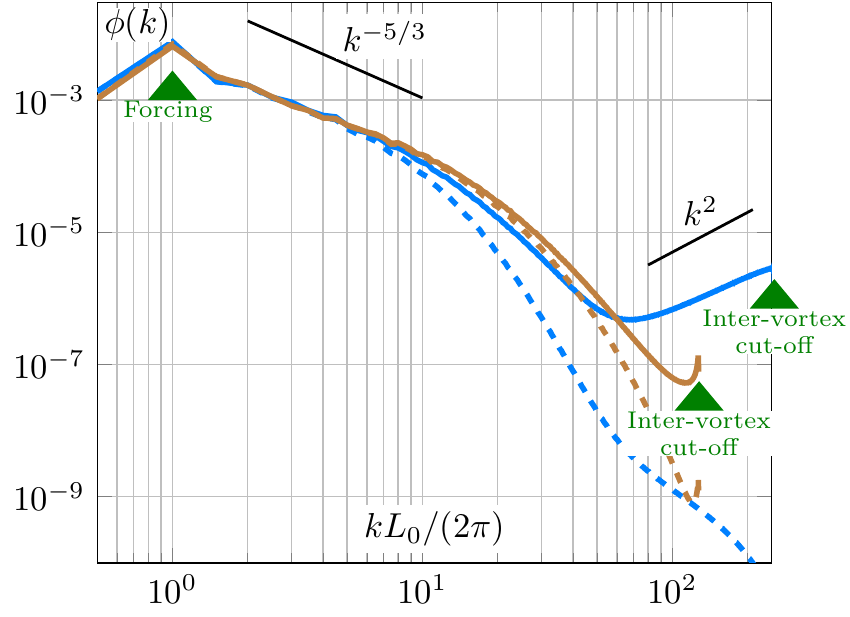}
\end{center}
\vspace*{-5mm}
\caption{Simulated 3D velocity power spectra. Solid lines are obtained
  from the velocity field of the superfluid component,
  $\vec{v}_s$. Dashed lines are obtained from the velocity field of
  the normal component, $\vec{v}_n$. The sky blue spectra were obtained at
  very low temperature ($T=\SI{1.15}{K}$, $1024^3$) ; the chocolate
  spectra were obtained at high temperature ($T=\SI{2.1565}{K}$,
  $512^3$). The smallest resolved scale matches the inter-vortex
  spacing. $L_0$ is defined as the forcing scale.}\label{sim:spectres}
\end{figure}

The velocity power spectra for normal and superfluid components are
displayed in figure \ref{sim:spectres} in the very-low-temperature and high-temperature limits: $T=\SI{1.15}{K}$ and $T=\SI{2.1565}{K}$
corresponding  to $\rho_s/\rho_n = 40$ and $\rho_s/\rho_s =
0.1$, respectively.  In order to allow closer comparisons with experiments,
the Reynolds number $Re$ is estimated as
\begin{equation}
Re = \frac{L_0\sqrt{\left<v_m^2\right>}}{\mu/\rho}
\end{equation}
where $v_m = \left(\rho_nv_n+\rho_sv_s\right)/{\rho}$ is the
momentum velocity\footnote{We used the one-dimensional rms value,
  $v_{\text{rms,1d}} = {v_{\text{rms,3d}}}/{\sqrt{3}}$ to be
  comparable with experiments.}, $L_0=\pi$ is the length-scale corresponding
to the forcing wave-number $k_0=1$ and $\mu/\rho$ is the kinematic viscosity.  The power spectrum of the momentum velocity is not
plotted but nearly matches the normal-component spectrum at high
temperature and the superfluid-component spectrum at very low
temperature, as expected from the mass density ratio.

The very-low-temperature  and high-temperature simulations have nearly the same Reynolds number: $Re=1960$
and $Re=2280$ respectively, which are much smaller than the Reynolds number of
the experiment: $Re = \num{1.8e5}$.  Yet, in both cases, the
spectra collapse at large scales close to a Kolmogorov's $k^{-5/3}$
scaling but differ at smaller scales, named
``meso-scales''\cite{salort2011}. In this range of meso-scales, larger
than the inter-vortex scale but smaller than inertial scales, the
superfluid component is no longer locked to the normal component. At
the lowest temperatures, its energy distribution approaches a $k^2$
scaling, as evidenced in figure \ref{sim:spectres}, which is
compatible with the equipartition of superfluid energy.

\begin{figure}[htbp]
\begin{center}
\includegraphics[scale=.90]{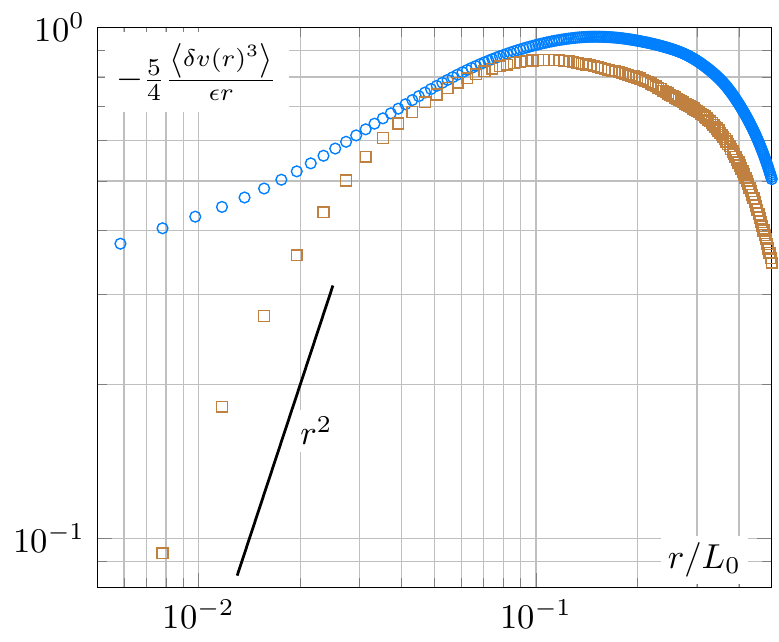}
\end{center}
\vspace*{-5mm}
\caption{Compensated third-order structure function obtained in
  numerical simulations at high temperature (chocolate squares) and very low
  temperature (sky blue circles) for nearly the same Reynolds
  numbers.}\label{fig:howartsim}
\end{figure}

The momentum velocity third-order longitudinal structure function is estimated 
 by averaging the longitudinal increment along the three 
directions in one ``snapshot'' of the
flow\footnoteremember{Similar}{We obtain similar results if the
  velocity increments are computed with the velocity field from the
  dominant component rather than $v_m$. The momentum velocity is
  convenient because it is defined for all temperatures and comparable
  to what is measured in experiments.}. 
%
\color{black} One does not expect the 4/5-law to hold exactly at such moderate
Reynolds number,  discrepancies being related to the viscous dissipation (at small scales)
 and the external forcing (at large scales) \cite{antonia2006}.  
However, we observe at high temperature that (i) the compensated
third-order structure function reaches a maximum slightly lower than one, which is consistent with
reported observations in classical turbulence (at comparable Reynolds numbers)
\cite{ishihara2009}, and (ii) the
small-scale behavior goes typically like $r^2$ corresponding to the
continuous (or smooth) limit $\delta u(r) \sim r$. 
At very low temperature, the
velocity field is no longer smooth at very small scales. It exhibits
irregular fluctuations, down to the smallest scales, related to the
equipartition noise.  This yields a different behavior of $\left<
  \delta v(r)^3 \right>$ as shown in figure
\ref{fig:howartsim}. \color{black} 
It is important to mention that the (total) disssipation rate, $\epsilon$, is eventually a parameter of our simulations. 
Indeed, $\rho \epsilon$ equals the power of the external forces (by assuming stationarity). 
This injected power is fixed and kept constant in our numerical scheme \cite{salort2011}. 


\color{black} In the following, we address the departure from the {ideal} 4/5-law at small scales, \emph{i.e.} related 
to the viscous dissipation.
Let us mention that departure at large scales (related to the external forcing) is beyond the scope of the present study and 
does not spoil the present results.

\color{black} In classical
turbulence, the viscous dissipation is accounted in the Kármán-Howarth equation, which generalizes the 4/5-law at small
scales:
\begin{equation}
\left< \delta v(r)^3 \right> +\frac{4}{5}\epsilon r = 6\nu\frac{\mathrm{d}\left< \delta v(r)^2 \right>}{\mathrm{d}r}\label{eq:howarth}
\end{equation}

\color{black} This Kármán-Howarth equation can be interpreted as an exact 
scale-by-scale energy budget.  Physically, the right-hand side of Eq.\
\ref{eq:howarth} takes into account the energy that leaks out of the
cascade due to the viscous dissipation.  Such generalization applied to the two-fluid model 
contains a term associated with the mutual friction between the superfluid and normal
components, which can not be formulated (strictly speaking) into a form  similar to Eq.\
\ref{eq:howarth}.  However, we propose here to pursue an empirical approach and assess to what extent
the classical relation (Eq.\ \ref{eq:howarth}) can be applied to He~II. \color{black} Formally, an effective kinematic viscosity 
can be defined from the deviation to the 4/5-law at  
small scales. More precisely, let us introduce
\begin{equation}
\mathcal{N}(r) = \frac{\left< \delta v^3 \right> + \frac{4}{5}\epsilon r}{6\frac{\mathrm{d}\left< \delta v^2 \right>}{\mathrm{d}r}} \label{eq:nu}
\end{equation}
  
For a classical Navier-Stokes fluid, Eq.\ref{eq:howarth} implies that
$\mathcal{N}(r)$ should match the kinetic viscosity $\mu/\rho$ from the
``center'' of the inertial range down to the smallest scales.

\begin{figure}[htbp]
\begin{center}
\includegraphics[scale=.90]{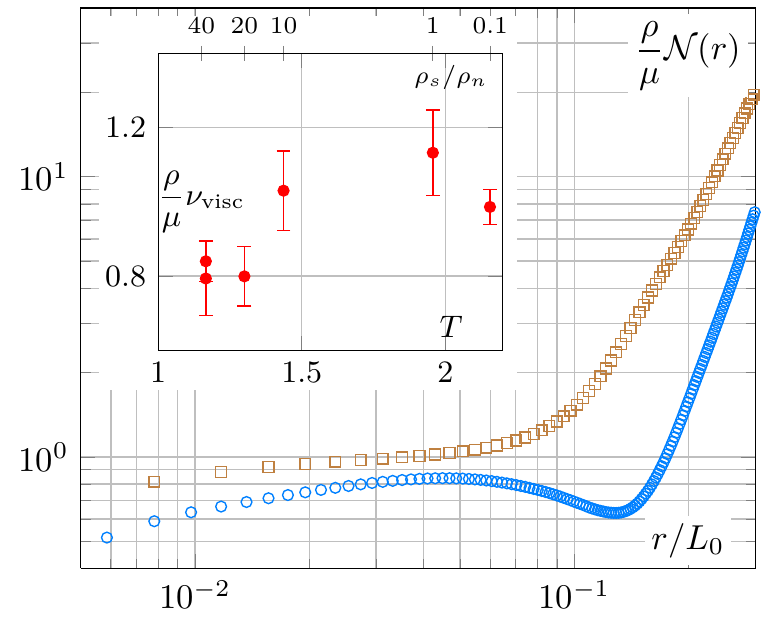}
\end{center}
\vspace*{-5mm}
\caption{Compensated effective viscosity versus scale
  obtained in numerical simulations at high temperature (chocolate squares)
  and very low temperature (sky blue circles) for nearly the same
  Reynolds number. Inset: effective viscosity estimated from the
  ``plateau'' of $\mathcal{N}(r)$ for various
  temperatures.} \label{fig:nueff}
\end{figure}

The values of $\mathcal{N}(r)$, normalized by $\mu/\rho$, 
are plotted versus scale in figure \ref{fig:nueff}.  
For all simulated
temperatures, ranging from $T=\SI{1.15}{K}$ ($\rho_s/\rho_n = 40$)  to $T= \SI{2.1565}{K}$
($\rho_s/\rho_n = 0.1$), this
plot exhibits a ``plateau'' in the inertial range, quite analogous to what is expected for a classical fluid. 
This means that the deviation to the 4/5-law can
be described (at first approximation) by introducing a constant effective viscosity. 
Interestingly, this remains valid even at very low temperatures, where the density of the normal component is also very small.  
This implies that the mutual friction term in the
superfluid equation (Eq.\ \ref{eq:vs}) cannot be neglected at very low temperature (even if it is proportional to
$\rho_n/\rho \ll 1$) and that it mimics to some extent a ``viscous leak'' along the
cascade. Nevertheless, $\mathcal{N}(r)$ deviates from the plateau at
the smallest scales, where both components are no longer locked,
especially at very low temperature (sky blue circles). This is in contrast
with classical turbulence, for which the ``plateau'' would extend down to
the smallest scales \cite{moisy1999}.

From the ``plateau''-value of $\mathcal{N}(r)$, we define the effective viscosity
$\nu_{\text{visc}}$. The estimates of $\nu_{\text{visc}}$ (compensated by $\mu/\rho$) for
various temperature and Reynolds-number conditions are gathered in the inset of
figure \ref{fig:nueff}. It is remarkable
that this effective viscosity matches the dynamic viscosity of the
normal component (normalized by the total density) within
\SI{20}{\percent} for all temperatures. As a result, these simulations indicate
that superfluid helium (He~II) behaves roughly as a viscous fluid at scales for which both normal and superfluid components are
nearly locked, \emph{i.e.} along the energy cascade. Furthermore, this feature remains satisfied  at the lowest temperatures, where the normal (viscous)
component fraction is smaller than \SI{3}{\percent}.

\section{Concluding remarks}

Using  third-order longitudinal velocity structure functions, we have argued both
experimentally and numerically that (stationary) turbulence in superfluid helium is consistent with an
energy cascade in the sense of Kolmogorov's theory. In
particular, our \color{black} experimental \color{black} data are
quantitatively compatible with the classical 4/5-law in the inertial
range.  It is worth pointing out that structure functions have been 
analyzed in the usual way because vortex singularities of the superfluid have been
smoothed out, either by the large-size (compared with the inter-vortex distance) probe or by the coarse-grained resolution of 
the simulation model. Without this low-pass filtering of the details of the
superfluid vortex tangle, comparisons with classical turbulence would have been less
straightforward.

The ``energy leak'' from the cascade was assessed by applying the
Kármán-Howarth equation on simulated velocity fields. We
find that He~II behaves as a viscous fluid in its cascade range with
an effective viscosity, $\nu_{\text{visc}}$, inherited from the normal
component, even down to the lowest temperature
($\rho_s/\rho_n=40$). This conclusion does not extent down to the
smallest (meso)-scales when both components are unlocked and
quasi-equipartition is evidenced.  It is interesting to compare
$\nu_{\text{visc}}$ with an (other) effective viscosity, $\nu_{\text{eff}}$,
 defined in the literature as
\cite{vinen2002}
\begin{equation}
\epsilon = \nu_{\text{eff}}\left( \frac{\kappa}{\delta^2} \right)^2 \simeq \nu_{\text{eff}}\left| \omega_s \right|^2 \label{eq:usualdef}
\end{equation}
These two  viscosities are comparable at
high temperature \cite{niemela2005}, which can be understood by writing
that both normal and superfluid components are roughly locked down to the (viscous)
dissipation length scale:
\begin{equation}
\nu_{\text{eff}} \equiv \epsilon\left|\omega_s\right|^{-2} \simeq \epsilon\left|\omega_n\right|^{-2} = \frac{\mu}{\rho_n} \simeq \frac{\mu}{\rho} = \nu_{\text{visc}}
\end{equation}

However, $\nu_{\text{eff}}$ departs from the ``viscous viscosity'',
$\nu_{\text{visc}}$, as the temperature is lowered
\cite{niemela2005,chagovets2007,walmsley2008}, but becomes compatible
with the ``friction viscosity'', $\nu_{\text{frict}} =
\kappa\frac{\rho_nB}{2\rho}$. This latter viscosity can be derived
from Eqs.\ \ref{eq:vn} and \ref{eq:vs} assuming that both components
are unlocked at small scales, which entails dissipation by friction of
one fluid component on the other \cite{roche2009} (see
\cite{vinen2002} for a microscopic derivation).  Thus, the definition
of $\nu_{\text{eff}}$ encompasses the two dissipative mechanisms 
occuring in He~II at finite temperature ($T \geq \SI{1}{K}$), \emph{i.e.} the ``viscous
dissipation'', $\nu_{\text{visc}}$, that we discuss in this letter, and
the ``friction dissipation'', $\nu_{\text{frict}}$. It would be
interesting to  understand  how 
$\nu_{\text{eff}}$ (Eq.\ \ref{eq:usualdef}) depends on the relative
weight of the two dissipation mechanisms and on a third dissipation
mechanism relevant in the zero temperature limit: sound emission by
vortex line \cite{nore1997_prl,vinen2000_prb,leadbeater2001}. \color{black} The
analytical integration of the Kármán-Howarth for the two-fluid model, which
implies additional modeling, would open this perspective. \color{black}

As a perspective to further understand the mechanisms leading to
viscous-like behavior, we point out a possible analogy with
classical truncated Euler systems, in which the presence of an
equipartitioned reservoir at small scales acts as a
molecular viscosity at larger scales
\cite{kraichnan1989,cichowlas2005,bos2006}.

\section{Acknowledgments}

This work benefited from the 
support of ANR 
(ANR-09-BLAN-0094) and from the computing facilities of PSMN at ENS
Lyon and of GENCI-CINES (grant 2011-026380).  We are grateful to
Grégory Garde who designed and built the Helium wind tunnel and to
Pierre Chanthib, Étienne Ghiringhelli, Pierre-Luc Delafin, Jacques
Depont and Jean-Luc Kueny for their help. We thank Laurent Chevillard,
Yves Gagne, Bernard Castaing and Roberto Benzi for interesting
discussions.

\bibliographystyle{eplbib}

\begin{thebibliography}{10}
\expandafter\ifx\csname url\endcsname\relax\def\url#1{\texttt{#1}}\fi

\footnotesize{ 

\bibitem{kolmogorov1941}
\Name{Kolmogorov A.} \REVIEW{C. R. Acad. Sci. USSR }{30}{1941}{301-305}.

\bibitem{landau1941}
\Name{Landau L.} \REVIEW{Phys. Rev. }{60}{1941}{356}.

\bibitem{vinen2002}
\Name{Vinen W.~F. \and Niemela J.~J.} \REVIEW{J. Low Temp. Phys.
  }{128}{2002}{167}.

\bibitem{sergeev2011}
\Name{Sergeev Y.} \REVIEW{Nature Physics }{7}{2011}{451}.

\bibitem{roche2007}
\Name{Roche P.-E. \emph{et al.}} \REVIEW{EPL }{77}{2007}{66002}.

\bibitem{bradley2008}
\Name{Bradley D. \emph{et al.}} \REVIEW{Phys. Rev. Lett. }{101}{2008}{065302}.

\bibitem{stalp1999}
\Name{Stalp S.~R., Skrbek L. \and Donnelly R.~J.} \REVIEW{Phys. Rev. Lett.
  }{82}{1999}{4831}.

\bibitem{niemela2005}
\Name{Niemela J., Sreenivasan K. \and Donnelly R.} \REVIEW{J. Low Temp. Phys.
  }{138}{2005}{537}.

\bibitem{chagovets2007}
\Name{Chagovets T., Gordeev A. \and Skrbek L.} \REVIEW{Phys. Rev. E
  }{76}{2007}{027301}.

\bibitem{walmsley2008}
\Name{Walmsley P. \and Golov A.} \REVIEW{Phys. Rev. Lett. }{100}{2008}{245301}.

\bibitem{rousset1994}
\Name{Rousset B. \emph{et al.}} in proc. of \Book{15\textsuperscript{th} Int. Cryo. Eng. Conf.}
Cryogenics,  Vol.~34 Supplement 1, 1994 pp. 317--320.

\bibitem{smith1999}
\Name{Smith M.~R., Hilton D.~K. \and Sciver S. W.~V.} \REVIEW{Phys. fluids
  }{11}{1999}{751}.

\bibitem{fuzier2001}
\Name{Fuzier S. \emph{et al.}} \REVIEW{Cryogenics
  }{41}{2001}{453}.

\bibitem{maurer1998}
\Name{Maurer J. \and Tabeling P.} \REVIEW{EPL }{43}{1998}{29}.

\bibitem{salort2010}
\Name{Salort J. \emph{et al.}} \REVIEW{Phys. fluids
  }{22}{2010}{125102}.

\bibitem{benzi2011}
\Name{Benzi R.} in \Book{Classical and Quantum Turbulence Workshop, Abu Dhabi},
  May 2\textsuperscript{nd}, 2011.

\bibitem{samuels1999}
\Name{Samuels D.~C. \and Kivotides D.} \REVIEW{Phys. Rev. Lett.
  }{83}{1999}{5306}.

\bibitem{donnelly1998}
\Name{Donnelly R. \and Barenghi C.} \REVIEW{J. Phys. Chem. Ref. Data
  }{27}{1998}{1217}.

\bibitem{kivotides2002} 
\Name{Kivotides D. \emph{et al.}} \REVIEW{EPL }{57}{2002}{845}.

\bibitem{salort2011}
\Name{Salort J. \emph{et al.}} \REVIEW{EPL
  }{94}{2011}{24001}.

\bibitem{pinton1994}
\Name{Pinton J.-F. \and Labb{\'e} R.} \REVIEW{J. Phys. II }{4}{1994}{1461}.

\bibitem{monin1971}
\Name{Monin A. \and Yaglom A.} \Book{Statistical Fluid Mechanics} (MIT Press,
  Cambridge) 1971.

\bibitem{chevillard2006}
\Name{Chevillard L. \emph{et al.}}
  \REVIEW{Physica D }{218}{2006}{77}.

\bibitem{antonia2006}
\color{black} \Name{Antonia R.~A. \and Burattini P.} \REVIEW{J. Fluid Mech.
  }{550}{2006}{175}. \color{black}

\bibitem{moisy1999}
\Name{Moisy F. \emph{et al.}} \REVIEW{Phys. Rev. Lett.
  }{82}{1999}{3994}.

\bibitem{ishihara2009}
\Name{Ishihara T. \emph{et al.}} \REVIEW{Annu. Rev. Fluid Mech.
  }{41}{2009}{165}.

\bibitem{barenghi1983}
\Name{Barenghi C. \and Donnelly R.} \REVIEW{J. Low Temp. Phys.
  }{52}{1983}{189}.

\bibitem{roche2009}
\Name{Roche P.-E. \emph{et al.}} \REVIEW{EPL
  }{87}{2009}{54006}.

\bibitem{nore1997_prl}
\Name{Nore C. \emph{et al.}} \REVIEW{Phys. Rev. Lett.}{78}{1997}{3896}.

\bibitem{vinen2000_prb}
\Name{Vinen W.} \REVIEW{Phys. Rev. B }{61}{2000}{1410}.

\bibitem{leadbeater2001}
\Name{Leadbeater M. \emph{et al.}}\REVIEW{Phys. Rev. Lett. }{86}{2001}{1410}.

\bibitem{kraichnan1989}
\Name{Kraichnan R. \and Chen S.} \REVIEW{Physica D }{37}{1989}{160}.

\bibitem{cichowlas2005}
\Name{Cichowlas C. \emph{et al.}} \REVIEW{Phys.
  Rev. Lett. }{95}{2005}{264502}.

\bibitem{bos2006}
\Name{Bos W. \and Bertoglio J.-P.} \REVIEW{Phys. fluids }{18}{2006}{071701}.

}

\end{thebibliography}



\end{document}